\documentclass{article}
\usepackage[utf8]{inputenc}
\usepackage{hyperref, amsmath, amssymb, tensor, bm, comment}
\usepackage{natbib}
\usepackage{todonotes}
\usepackage{enumitem, setspace}

\renewcommand{\-}[1]{\overline{#1}}

\title{Conformal Invariance of the Newtonian Weyl Tensor}
\author{Neil Dewar\footnote{neil.dewar@lrz.uni-muenchen.de}~ and James Read\footnote{james.read@pmb.ox.ac.uk (Corresponding author.)}}
\date{}
\begin{document}

\maketitle

\begin{abstract}
    It is well-known that the conformal structure of a relativistic spacetime is of profound physical and conceptual interest. In this note, we consider the analogous structure for Newtonian theories. We show that the Newtonian Weyl tensor is an invariant of this structure.
\end{abstract}




\section{Conformal Leibnizian spacetimes}\label{s5.1}

We begin by introducing a \emph{Leibnizian spacetime}, which is a triple $\left(  M, t_a , h^{ab} \right)$, where (i) $M$ is a differentiable manifold; (ii) $t_a$ is a non-vanishing, closed 1-form; and (iii) $h^{ab}$ is a positive semidefinite symmetric tensor such that $h^{ab} t_b = 0$. A connection $\nabla$ on $M$ is said to be \emph{compatible} with this spacetime if and only if
\begin{subequations}
\begin{align}
    \nabla_a t_b &= 0,\\
    \nabla_a h^{bc} &= 0.
\end{align}
\end{subequations}
We will confine our attention to spacetimes which are \emph{spatially flat}: that is, which are such that the Riemann tensor $R\indices{^a_{bcd}}$ of any compatible connection obeys $h^{rb} h^{sc} h^{td} R\indices{^a_{bcd}} = 0$. (One can show that if this holds of any one compatible connection, it holds of all of them.)

Because of the separation of the spatial and temporal metrical structure, we have scope to vary conformally the spatial and temporal structure independently of one another (although as we shall see, there are reasons to couple the two kinds of conformal transformation). Consider, first, a conformal transformation of the temporal structure
\begin{equation}
    t_a \mapsto \xi^2 t_a ,
\end{equation}
where $\xi$ is a nowhere-vanishing \emph{and spatially constant} scalar field. To say that $\xi$ is spatially constant means that $h^{ab} d_b \xi = 0$. This is equivalent to ensuring that the conformally transformed temporal 1-form is still closed and thus that there exists a global time function (and so a notion of Newtonian absolute time) in the conformally-transformed model.\footnote{Cf.~\cite[ch.~4]{Malament}, \cite{BM}. Note that throughout this work we assume that the manifold $M$ is simply connected.} If we replace the temporal 1-form in a Leibnizian spacetime with a conformal equivalence class thereof, we obtain \emph{Machian spacetime}.

Second, consider a conformal transformation of the spatial structure,
\begin{equation}
    h^{ab} \mapsto \lambda^2 h^{ab},
\end{equation}
where $\lambda$ is, again, a nowhere-vanishing and spatially constant scalar field. This time, we require that $\lambda$ be spatially constant in order to preserve spatial flatness of the spacelike hypersurfaces. If we replace the spatial metric in a Leibnizian spacetime with a conformal equivalence class of spatial metrics, then we obtain \emph{spatially conformal Leibnizian spacetime}.

Finally, we may consider joint conformal transformations of the spatial and temporal structure:
\begin{subequations}
\begin{align}
    t_a &\mapsto \frac{1}{\lambda^2} t_a , \label{t-conf}\\
    h^{ab} &\mapsto \lambda^2 h^{ab} . \label{h-conf}
\end{align}
\end{subequations}
where $\lambda$ is a nowhere-vanishing and spatially constant scalar field. As we will show in the next section, it is conformal transformations of this kind which preserve the Newtonian analogue of the Weyl tensor. A spacetime equipped with a conformal equivalence class of $(t_a, h^{ab})$ pairs will be referred to as a \emph{conformal Leibnizian spacetime}.\footnote{For definitions of Newtonian conformal structure complimentary to our own, see \cite{Curiel, ES, Duval2, Duval}. The former two of these papers define a notion of Newtonian conformal structure in order to generalise the constructive axiomatics of \cite{EPS} to the case of Newton-Cartan theory.}

\section{Invariance of the Newtonian Weyl tensor}\label{s5.2}

Consider a relativistic spacetime $\left( M , g_{ab} \right)$. From $g_{ab}$ and its associated Levi-Civita derivative operator, one can define the \emph{Weyl tensor} of this spacetime, which is the trace-free part of the Riemann tensor:\footnote{For the generalisation to arbitrary spacetime dimensions, see e.g.~\cite[p.~40]{Wald}.}
\begin{equation}
    C\indices{^{a}_{bcd}} = R\indices{^{a}_{bcd}} - \frac{1}{2}\left( \delta\indices{^{a}_{[d}} R\indices{_{c]b}} + g\indices{_{b[c}}R\indices{_{d]}^{a}} \right) - \frac{1}{3} R \delta\indices{^{a}_{[c}}g\indices{_{d]b}}.
\end{equation}
This object is invariant under conformal transformations of $g_{ab}$; thus, it will be the same for all points in the affine space of connections compatible with a given conformal structure.

Now consider a Leibnizian spacetime endowed with a Newtonian connection. At \cite[p.~574]{DW}, the authors proposed the following Newtonian analogue of the Weyl tensor:\footnote{The generalisation to arbitrary spacetime dimensions is straightforward:~one replaces the denominator of the second term on the right hand side with $\left( n-1\right)$.}
\begin{equation}\label{W1}
C\indices{^{a}_{bcd}} = R\indices{^{a}_{bcd}} - \frac{2}{3} \delta\indices{^{a}_{[d}} R\indices{_{c]b}}.
\end{equation}
\cite{DW} were not the first to construct a Newtonian Weyl tensor---\cite{Ehlers} apply `frame theory' (a unified framework for both relativistic and classical spacetimes\footnote{\cite{Fletcher} claims that a topology can be introduced on the space of solutions of frame theory, such that Newton-Cartan theory can be understood as the non-relativistic limit of general relativity; this (he claims) affords a precise sense in which Newton-Cartan theory can be reduced to general relativity. Though we concur with these results, we wish to flag that there are other senses of the reduction of Newton-Cartan theory to general relativity which do not involve taking limits---for example null reduction, in which Newton-Cartan theory is directly embedded into (certain solutions of) five-dimensional general relativity.}) in order to take the non-relativistic limit of the general relativity Weyl tensor; the result is:\footnote{Another definition of the Newtonian Weyl tensor, equivalent to that of \cite{DW}, can be found in \cite{Duval}.}
\begin{equation}\label{W2}
C\indices{^{a}_{bcd}} = R\indices{^{a}_{bcd}} - \frac{8\pi G \rho}{3} t_b \delta\indices{^{a}_{[c}}t\indices{_{d]}}.
\end{equation}
On-shell in Newton-Cartan theory---so that the geometrised Poisson equation
\begin{equation}\label{Poisson}
R_{ab}=4\pi G \rho t_a t_b
\end{equation}
holds---\eqref{W1} is identical to \eqref{W2}. This gives us confidence that \eqref{W1} is indeed the correct object to represent a Newtonian Weyl tensor. We should flag, though, that it is not obviously appropriate to use the on-shell version of the Weyl tensor \eqref{W2}, for the Poisson equation is \emph{not} invariant under conformal rescalings (just as the Einstein equation in general relativity is not invariant under conformal rescalings of the metric field $g_{ab}$).\footnote{To see this, one need only take \eqref{Poisson}, conformally transform both $t_a$ and $h^{ab}$, and substitute for $R_{ab}$ with \eqref{R-transfo}, where $U\indices{^{a}_{bc}}$ is the difference tensor associated with the derivative operator compatible with $t_a$ and $h^{ab}$, and the derivative operator compatible with their conformally transformed versions (see below). Note, in particular, that \eqref{Poisson} is not invariant even under the specific class of conformal transformations given by \eqref{t-conf} and \eqref{h-conf}.} Thus, in the remainder we focus upon the version of the Weyl tensor \eqref{W1}---our goal now is to show that this object is invariant under conformal rescalings of $t_a$ and $h^{ab}$, and thus is (one might say) a gauge-invariant quantity in any theory set in a conformal Leibnizian spacetime. One further benefit of using \eqref{W1} rather than \eqref{W2} is that we do not commit ourselves to working with the dynamics of Newton-Cartan theory.


We now show that this object, in analogy with the Newtonian case, is invariant under (an important class of) conformal rescalings of $t_{a}$ and $h^{ab}$. We begin with a spatially flat classical spacetime $(M, t_a, h^{ab}, \nabla)$, where $M$ is simply connected and $\nabla$ satisfies the curvature condition
\begin{equation}
    R\indices{^a_b^c_d} = R\indices{^c_d^a_b} .
\end{equation}
In light of these facts,\footnote{\cite[Proposition 4.3.7]{Malament}} we may introduce an observer field $N^a$: a unit timelike field which is geodesic and twist-free with respect to $\nabla$, i.e. which satisfies
\begin{subequations}
\begin{align}
    N^a \nabla_a N^b &= 0,\\
    h^{ab} \nabla_b N^c &= h^{cb} \nabla_b N^a.
\end{align}
\end{subequations}
Relative to this field, we may introduce a spatial metric $h_{ab}$, which is the unique symmetric field satisfying the conditions
\begin{subequations}
\begin{align}
    h_{ab} N^a &= 0,\\
    h^{ab} h_{bc} &= \delta^a_c - N^a t_c.
\end{align}
\end{subequations}

Now suppose we apply the conformal transformations\footnote{\label{fn-blah}These are not quite the transformations one would have expected: purely on dimensional grounds, one might have expected that $h^{ab} \mapsto \lambda^4 h^{ab}$ (given $t_a \mapsto \lambda^{-2} t_a$). Our reason for using the transformations presented in the main text is simply that this choice yields invariance of the Newtonian Weyl tensor; unfortunately, we don't have a good explanation of why invariance is guaranteed by this choice, rather than by the more natural one. (Thanks to Jim Weatherall for raising this concern.)

It does bear mentioning, however, that the above two `expected' transformations yield invariance of the Newtonian Weyl tensor, \emph{without imposition of the restriction of spatial constancy}, but at the price that the transformed derivative operator compatible with the rescaled $h^{ab}$ and $t_a$ need not be torsion-free. If one desires the conformal transformations retain torsion-freeness, then one must choose the transformations discussed in the main body of this text, and also impose the spatial constancy condition (which guarantees spatial flatness).}
\begin{subequations}
\begin{align}
    t_a &\mapsto \-t_a = \frac{1}{\lambda^2} t_a,\\
    h^{ab} &\mapsto \-h^{ab} = \lambda^2 h^{ab}.
\end{align}
\end{subequations}
where $\lambda$ is a spatially constant, nowhere-vanishing scalar field: $h^{ab} \nabla_b \lambda = 0$. It follows\footnote{\cite[Proposition 4.1.1]{Malament}} that there is a scalar field $\kappa$ such that $\nabla_a \lambda = \kappa t_a$; explicitly,
\begin{equation}
    \kappa = N^a \nabla_a \lambda.
\end{equation}
Note that $\kappa$, too, is spatially constant: for, $\nabla_n \kappa = \nabla_n (N^a \nabla_a \lambda) = t_n (N^a \nabla_a \kappa)$. We will use this observation below.

We now wish to find the `conformally transformed' version of $\nabla$. Unlike in the relativistic case, we do not obtain such a transformed connection merely from having transformed the temporal and spatial metrics, since they do not uniquely determine the connection. However, the metrics together with a unit timelike field \emph{do} uniquely determine a connection: namely, the unique connection with respect to which the timelike field is an observer field (i.e., is geodesic and twist-free).\footnote{\cite[Proposition 4.3.4]{Malament}} We therefore define\footnote{This condition can be interpreted as a conformal transformation of the derivative operator $\nabla$---cf.~\cite[\S4.2]{Duval}.} 
\begin{equation}
    \-N^{a} := \lambda^2 N^a ,
\end{equation}
which is a unit timelike field relative to $\-t_a$. We then define $\-\nabla := (\nabla, U\indices{^a_{bc}})$, where
\begin{equation}\label{U}
    U\indices{^a_{bc}} := \frac{2\kappa}{\lambda} t\indices{_{(b}} \delta\indices{^a_{c)}} .
\end{equation}
Some straightforward computations verify that $\-\nabla$ is compatible with $\-t_a$ and $\-h^{ab}$, and that $\-N^a$ is geodesic and twist-free with respect to $\-\nabla$. Since $U\indices{^a_{bc}}$ is independent of $N^a$, we may indeed regard $\-\nabla$ as \emph{the} conformally transformed version of $\nabla$: had we chosen to represent $\nabla$ via a different observer field ${N'}^a$, we would nevertheless have obtained the same $U\indices{^a_{bc}}$, and hence the same $\-\nabla$.

Next, recall that\footnote{\cite[Problem 1.8.1]{Malament}}
\begin{equation}\label{R-transfo}
    \-R\indices{^a_{bcd}} = R\indices{^a_{bcd}} + 2 \nabla\indices{_{[c}} U\indices{^a_{d]b}} + 2 U\indices{^n_{b[c}} U\indices{^a_{d]n}}.
\end{equation}
Plugging in \eqref{U}, we obtain
\begin{equation}
    \-R\indices{^a_{bcd}} = R\indices{^a_{bcd}} + 2 \left(\frac{N^n \nabla_n \kappa}{\lambda} \right) t_b t\indices{_{[c}} \delta\indices{^a_{d]}}.
\end{equation}
Note that it follows from this that $\-R^{abcd} = 0$, i.e. the conformally transformed spacetime is spatially flat (given that the original spacetime was spatially flat). From here, it is easy to compute the Ricci tensor $\-R_{bc} = \-R\indices{^a_{bca}}$:\footnote{Note that here the number of dimensions becomes relevant, since we have used the fact that in four dimensions, $\delta\indices{^a_a} = 4$.}
\begin{equation}
    \-R_{bc} = R_{bc} + 3 \left(\frac{N^n \nabla_n \kappa}{\lambda}\right) t_b t_c
\end{equation}
It remains only to substitute these expressions into \eqref{W1}, from which we obtain
\begin{equation}
    \-C\indices{^a_{bcd}} = C\indices{^a_{bcd}}.
\end{equation}
I.e., the Newtonian Weyl tensor, like its relativistic cousin, is invariant under these conformal transformations. This nuances a suggestion in \cite[p.~573]{DW} that this object is not conformally invariant, and also the subsequent suggestion that ``conformal transformations just do not have any physical significance in geometrized Newtonian gravitation''---what we find is that, under a certain \emph{class} of conformal transformations (namely, those which are spatially constant), the Newtonian Weyl tensor \emph{is} conformally invariant.\footnote{Cf.~also footnote \ref{fn-blah}.}


Finally, we note that since the symmetries of the Riemann tensor are the same as those of the Levi-Civita connection in the relativistic case \cite[p.~258]{Malament}, we also expect the Newtonian Weyl tensor to vanish identically in spacetime dimensions $D\leq 3$.

\section{A degeometrised Weyl tensor}

Newton-Cartan theory and Newtonian gravitation theory are related via the Trautman geometrisation and recovery theorems \cite[ch.~4]{Malament}. \cite{Ellis} remarks that the degeometrised Newtonian analogue of the Riemann tensor is $\nabla^i \nabla^j \phi$, and that the trace-free part of this object is (here, we follow Ellis in using spatial indices)\footnote{See \cite{Buchert} and \cite{WallaceNG} for further discussion of this object, sometimes called the \emph{Newtonian tidal tensor}.}
\begin{equation}\label{eqE}
    E^{ij} := \nabla^i \nabla^j \phi - \frac{1}{3} h^{ij} \nabla_k \nabla^k \phi .
\end{equation}
A result with the same structural form as Ellis' $E^{ij}$ can be derived directly, and in a coordinate-independent way, using Trautman recovery. First, recall that a Newton-Cartan connection $\tilde{\nabla}$ is related to a degeometrised Newtonian connection $\nabla$ via $\tilde{\nabla} = \left( \nabla, C\indices{^{a}_{bc}}\right)$, where $C\indices{^{a}_{bc}} = -t_b t_c \nabla^a \phi$. The Riemann and Ricci tensors for the Newton-Cartan connection can then be written in terms of the degeometrised gravitational potential $\phi$, as\footnote{\cite[pp.~268-269]{Malament}.}
\begin{subequations}
\begin{align}
    \tilde{R}\indices{^{a}_{bcd}} &= -2 t_b t\indices{_{[d}} \nabla\indices{_{c]}} \nabla^a \phi \label{Riem1}, \\
    \tilde{R}_{bc} &= t_b t_c \nabla_n \nabla^n \phi . \label{Riem2}
\end{align}
\end{subequations}
One can then substitute \eqref{Riem1} and \eqref{Riem2} into \eqref{W1} in order to express the Weyl tensor in terms of $\phi$; one finds:
\begin{equation}
    \tilde{C}\indices{^{a}_{bcd}} = -2 t_b t\indices{_{[d}} \nabla\indices{_{c]}} \nabla^a \phi - \frac{2}{3} \delta\indices{^{a}_{[d}}t\indices{_{c]}} t_b \nabla_n \nabla^n \phi.
\end{equation}
This is the four-dimensional analogue of Ellis' object. As shown by \cite{Ehlers}, \eqref{eqE} is the `electric' part of the Newtonian Weyl tensor (when expressed in terms of the gravitational potential $\phi$).

\section{Applications}
There remains much work to be done with the Newtonian Weyl tensor. For example:
\begin{enumerate}
    \item Demonstrate that the Newtonian Weyl tensor is appropriately related to the Newtonian analogues of e.g.~the Schouten, Lanczos, and Plebanski tensors.
    \item Use the Newtonian Weyl tensor to construct a non-relativistic analogue of the Petrov classification.\footnote{See e.g.~\cite[p.~179]{Wald}.}
    \item Use the Newtonian Weyl tensor to explore gravitational waves in Newton-Cartan theory.\footnote{The status of gravitational waves in Newtonian theories is a matter of ongoing debate: \cite[p.~6]{Hansen} claim that these do not exist, while \cite[p.~574]{DW} demur. In \cite{LR}, the authors explicitly construct (albeit using unorthodox compositions of standing waves) propagating wave solutions in Newton-Cartan theory, and associate these with a non-vanishing Weyl tensor. Although there is certainly more to be done on these fronts (for example, one might reasonably question that naturalness of the constructions of \cite{LR}), it is exactly this kind of work which we have in mind when we make this point.}
    \item Use the conformal Newtonian spacetimes to write e.g.~\emph{shape dynamics} in terms of fields on spacetime.\footnote{See \cite{Mercati} for an introduction to shape dynamics.}
\end{enumerate}
A more general moral of this work is the following. There are various geometrical sources of non-geodesic motion of test particles, in a given spacetime theory. One is torsion---as is well-known from the framework of teleparallel gravity (see e.g.~\cite{TPGbook}). In \cite{ReadTeh}, it was shown that Trautman recovery can be understood as a case of teleparallelisation; thus, the mechanism via which one can source non-geodesic motion in both Newtonian and relativistic theories by the introduction of torsion is exactly parallel. In this paper, we have considered another potential source of non-geodesic motion:~the non-metricity naturally associated with conformal rescalings (see e.g.~\cite{Almeida}); again, we have shown that, technically, the introduction of such non-metricity into both contexts is parallel, for in both cases (e.g.)~the Weyl tensor is an invariant of the associated conformal structure. Thus, the structural aspects of both Newtonian and relativistic theories, once one introduces geometrical sources of non-geodesic motion such as torsion and non-metricity, are closely related.

\section{Acknowledgements}

We are very grateful to Erik Curiel, Dominic Dold, Sean Gryb, Nic Teh, Jim Weatherall, to the anonymous referees, and to the audience at the 2019 `ONB' spacetime workshop at Oxford, for helpful discussions and feedback.

\bibliography{Newton-Weyl}

\begin{thebibliography}{}

\bibitem[Aldrovandi and Pereira, 2013]{TPGbook}
Aldrovandi, R. and Pereira, J.~G. (2013).
\newblock {\em Teleparallel {{Gravity}}: {{An Introduction}}}.
\newblock {Springer Science \& Business Media}.

\bibitem[Almeida et~al., 2014]{Almeida}
Almeida, T.~S., Pucheu, M.~L., Romero, C., and Formiga, J.~B. (2014).
\newblock From {{Brans}}-{{Dicke}} gravity to a geometrical scalar-tensor
  theory.
\newblock {\em Physical Review D}, 89(6):064047.

\bibitem[Bekaert and Morand, 2016]{BM}
Bekaert, X. and Morand, K. (2016).
\newblock Connections and dynamical trajectories in generalised
  {{Newton}}-{{Cartan}} gravity {{I}}. {{An}} intrinsic view.
\newblock {\em Journal of Mathematical Physics}, 57(2):022507.

\bibitem[Buchert and Ostermann, 2012]{Buchert}
Buchert, T. and Ostermann, M. (2012).
\newblock Lagrangian theory of structure formation in relativistic cosmology:
  Lagrangian framework and definition of a nonperturbative approximation.
\newblock {\em Phyical Review D}, 86:023520.

\bibitem[Curiel, 2015]{Curiel}
Curiel, E. (2015).
\newblock A {W}eyl-type theorem for {G}eometrized {N}ewtonian {G}ravity.
\newblock {\em https://arxiv.org/abs/1510.02089}.

\bibitem[Dewar and Weatherall, 2018]{DW}
Dewar, N. and Weatherall, J.~O. (2018).
\newblock On {{Gravitational Energy}} in {{Newtonian Theories}}.
\newblock {\em Foundations of Physics}, 48(5):558--578.

\bibitem[Duval et~al., 2017]{Duval}
Duval, C., Gibbons, G., and Horv\'{a}thy, P. (2017).
\newblock Conformal and projective symmetries in {N}ewtonian cosmology.
\newblock {\em Journal of Geometry and Physics}, 112:197--209.

\bibitem[Duval and Horv\'{a}thy, 2009]{Duval2}
Duval, C. and Horv\'{a}thy, P. (2009).
\newblock Non-relativistic conformal symmetries and {N}ewton-{C}artan
  structures.
\newblock {\em Journal of Physics A:~Mathematical and Theoretical}, 42.

\bibitem[Ehlers and Buchert, 2009]{Ehlers}
Ehlers, J. and Buchert, T. (2009).
\newblock On the {{Newtonian}} limit of the {{Weyl}} tensor.
\newblock {\em General Relativity and Gravitation}, 41(9):2153--2158.

\bibitem[Ehlers et~al., 1972]{EPS}
Ehlers, J., Pirani, F., and Schild, A. (1972).
\newblock The geometry of free fall and light propagation.
\newblock {\em General Relativity: Papers in Honour of J.~L.~Synge}, pages
  63--84.

\bibitem[Ellis, 1971]{Ellis}
Ellis, G. (1971).
\newblock Relativistic cosmology.
\newblock {\em Proceedings of the international school of physics ``{E}nrico
  {F}ermi'', Course 47: General relativity and cosmology}, pages 104--182.

\bibitem[Ewen and Schmidt, 1989]{ES}
Ewen, H. and Schmidt, H.-J. (1989).
\newblock Geometry of free fall and simultaneity.
\newblock {\em Journal of Mathematical Physics}, 30:1480--1486.

\bibitem[Fletcher, 2019]{Fletcher}
Fletcher, S.~C. (2019).
\newblock On the {{Reduction}} of {{General Relativity}} to {{Newtonian
  Gravitation}}.
\newblock {\em Studies in History and Philosophy of Science Part B: Studies in
  History and Philosophy of Modern Physics}, 68:1--15.

\bibitem[Hansen et~al., 2019]{Hansen}
Hansen, D., Hartong, J., and Obers, N. (2019).
\newblock Gravity between {N}ewton and {E}instein.
\newblock {\em arXiv:1904.05706v2}.

\bibitem[Linnemann and Read, 2020]{LR}
Linnemann, N. and Read, J. (2020).
\newblock On the status of {N}ewtonian gravitational radiation.

\bibitem[Malament, 2012]{Malament}
Malament, D.~B. (2012).
\newblock {\em Topics in the Foundations of General Relativity and
  {{Newtonian}} Gravitation Theory}.
\newblock {University of Chicago Press}, {Chicago, IL}.

\bibitem[Mercati, 2018]{Mercati}
Mercati, F. (2018).
\newblock {\em Shape Dynamics: Relativity and Relationalism}.
\newblock {Oxford University Press}, {Oxford}.

\bibitem[Read and Teh, 2018]{ReadTeh}
Read, J. and Teh, N.~J. (2018).
\newblock The teleparallel equivalent of {{Newton}}\textendash{{Cartan}}
  gravity.
\newblock {\em Classical and Quantum Gravity}, 35(18):18LT01.

\bibitem[Wald, 1984]{Wald}
Wald, R. (1984).
\newblock {\em General Relativity}.
\newblock {University of Chicago Press}, {Chicago, IL}.

\bibitem[Wallace, 2017]{WallaceNG}
Wallace, D. (2017).
\newblock More problems for {N}ewtonian cosmology.
\newblock {\em Studies in History and Philosophy of Science Part B: Studies in
  History and Philosophy of Modern Physics}, 57:35--40.

\end{thebibliography}

\end{document}